\documentclass[floatfix,prl,twocolumn]{revtex4}
\usepackage{graphicx}
\usepackage{amsmath}
\usepackage{amssymb}
\usepackage{bm}

\DeclareMathOperator{\Real}{Re}
\DeclareMathOperator{\Li}{Li_{2}}

\begin{document}

\title{Toroidal Crystals}

\author{Luca Giomi}
%\email{lgiomi@physics.syr.edu}

\author{Mark J. Bowick}
%\email{bowick@physics.syr.edu}

\affiliation{Department of Physics,
Syracuse University,
Syracuse New York,
13244-1130}

\begin{abstract}
Crystalline assemblages of identical sub-units packed together and
elastically bent in the form of a torus have been found in the past
ten years in a variety of systems of surprisingly different nature,
such as viral capsids, self-assembled monolayers and carbon
nanomaterials. In this Letter we analyze the structural properties
of toroidal crystals and we provide a unified description based on
the elastic theory of defects in curved geometries. We find ground
states characterized by the presence of 5-fold disclinations on the
exterior of the torus and 7-fold disclinations in the interior. The
number of excess disclinations is controlled primarily by the aspect
ratio of the torus, suggesting a novel mechanism for creating
toroidal templates with precisely controlled valency via
functionalization of the defect sites.
\end{abstract}

\maketitle

\begin{figure}[t]
\centering
\includegraphics[width=1\columnwidth]{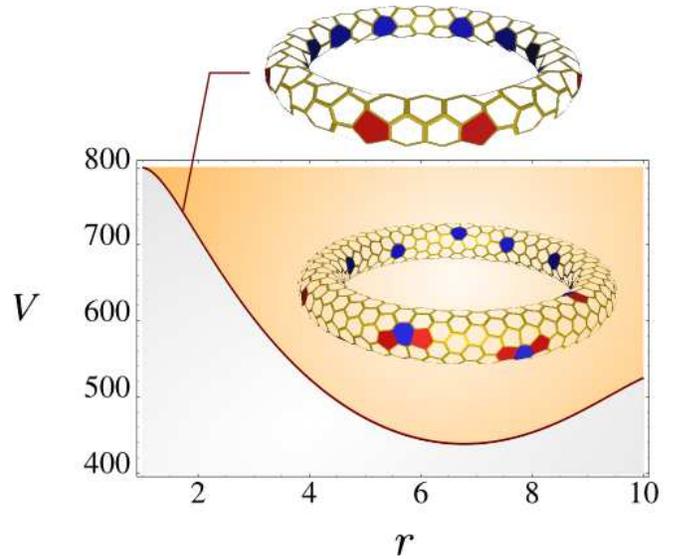}
\caption{\label{fig:phase_diagram3}(Color online) Isolated defects and 
scar phases in the $(r,V)$ plane. When the number of vertices $V$ 
increases the range of the screening curvature becomes smaller than 
one lattice spacing and disclinations appear delocalized in the form 
of a $5-7-5$ grain boundary mini-scar.}
\end{figure}

Toroidal micelles can be formed in the self-assembly of
dumbbell-shaped amphiphilic molecules \cite{KimEtAl:2006}. Molecular
dumbbells dissolved in a selective solvent self-assemble in
aggregate structures determined by their amphiphilic character. This
process yields coexisting spherical and open-ended cylindrical
micelles which evolve slowly over the course of a week to more
stable toroidal micelles. Toroidal geometries also occur in
microbiology in the viral capsid of the coronavirus \emph{torovirus}
\cite{SnijderHorzinek}. The torovirus is an RNA viral package of
maximal diameter between $120$ and $140$ nm and is surrounded, as
other coronaviridae, by a double wreath/ring of cladding proteins.

Carbon nanotori form another fascinating and technologically promising 
class of toroidal crystals \cite{LiuEtAl:1997} with remarkable magnetic 
and electronic properties. The interplay between the ballistic motion 
of the $\pi$ electrons and the geometry of the embedding torus leads 
to a \emph{colossal} paramagnetic moment in metallic carbon tori 
\cite{LiuEtAl:2002}. The high magnetic susceptibility, together with 
large ring radius (up to $0.03$ $\mu$m for single tubes of $1.4$ nm in 
diameter \cite{MartelSheaAvouris:1999}), implies that the mobility of 
the electrons in the highest occupied molecular orbital is several 
times larger than that of other ring-shaped molecules such as benzene
\cite{Haddon:1997}.

A unified theoretical framework to describe the structure of
toroidal crystals is provided by the elastic theory of defects in a
curved background \cite{PerezGarridoEtAl:1997,BowickNelsonTravesset:2000,
VitelliLucksNelson:2006,GiomiBowick:2007}. This formalism has the 
advantage of far fewer degrees of freedom than a direct treatment 
of the microscopic interactions and allows one to explore the origin 
of the emergent symmetry observed and expected in toroidal crystals 
as the result of the interplay between defects and geometry. Since 
defective regions are natural places for biological activity and 
chemical linking, a thorough understanding of the surface topology 
of crystalline assemblages could represent a significant step toward 
a first-principle design of entire libraries of nano and mesoscale 
components with precisely determined valency that could serve as the 
building blocks for mesomolecules or bulk materials via self-assembly 
or controlled fabrication.

The embedding of an equal number of pentagonal and heptagonal
disclination defects in a hexagonal carbon network was first
proposed by Dunlap in 1992 as a possible way to incorporate positive
and negative Gaussian curvature into the cylindrical geometry of
nanotubes \cite{Dunlap:1992}. In the Dunlap construction the
curvature is achieved by the insertion of ``knees'' in conjunction
with each pentagon-heptagon pair arising from the junction of
tubular segments of different chirality. The latter is
conventionally specified by two integers $(N,M)$ which identify the
direction along which a planar triangular lattice is rolled up in
the form of a tube \cite{SaitoEtAl:1998}. A junction between an
$(N,0)$ and an $(N,N)$ tube is obtained, for instance, by placing a
$7-$fold disclination along the internal equator of the torus and a
$5-$fold disclination along the external equator \cite{LambinEtAl:1995,
YaoEtAl:1999}. By repeating the $5-7$ connection periodically it is 
possible to construct an infinite number of toroidal lattices with 
an even number of disclinations pairs and the dihedral symmetry 
group $D_{nh}$ (where $2n$ is the total number of $5-7$ pairs).

Another class of crystalline toroids with dihedral antiprismatic
symmetry $D_{nd}$ was initially proposed by Itoh \emph{at al}
\cite{ItohEtAl:1993} shortly after Dunlap. Aimed at producing a
structure similar to C$_{60}$ fullerene, Itoh's original
construction implied ten disclination pairs and the symmetry group
$D_{5d}$. In contrast to Dunlap toroids, the disclinations in the
antiprismatic torus are not perfectly aligned along the equator
but rather staggered at some angular distance $\delta\psi$ from the
equatorial plane. A general classification scheme for $D_{nd}$
symmetric tori can be found in Ref. \onlinecite{BergerAvron:1995}.
In this paper we will refer to the lattices themselves, with $2n$
disclination pairs, by the symbols TP$n$ and TA$n$, with respective
symmetry groups $D_{nh}$ and $D_{nd}$.

Within the elastic theory of defects on curved surfaces 
\cite{PerezGarridoEtAl:1997,BowickNelsonTravesset:2000,
VitelliLucksNelson:2006,GiomiBowick:2007} the original interacting 
particle problem is mapped to a system of interacting disclination 
defects in a continuum elastic curved background. Disclinations are 
characterized by their topological or disclination charge, $q_{i}$, 
representing the departure of a vertex from a prefect triangular 
lattice. Thus $q_{i}=6-c_{i}$, where $c_{i}$ is the coordination 
number of the $i$th vertex. A classic theorem of Euler requires the 
total disclination charge of any triangulation of a two-dimensional 
manifold $M$ to be equal to $6\chi_{M}$, where $\chi_{M}$ is the 
Euler characteristic of $M$. In the case of the torus $\chi_{M}=0$, 
and thus disclinations must appear in pairs of opposite disclination 
charge (i.e. $5-$fold and $7-$fold vertices with $q_{i}=1$ and $-1$ 
respectively) in order to ensure disclination charge neutrality. 
The total free energy of a toroidal crystal with $N$ disclinations 
can be expressed as
\begin{equation}\label{eq:free_energy}
F_{el} = \frac{1}{2Y}\int_{M} d^{2}x\,\Gamma^{2}(\bm{x})+\epsilon_{c}\sum_{i=1}^{N}q_{i}^{2}+F_{0}\,,
\end{equation}
where $Y$ is the 2D-Young modulus and $\Gamma(\bm{x})$ is the
solution of the following Poisson problem with periodic boundary
conditions:
\begin{equation}\label{eq:poisson}
\Delta\Gamma(\bm{x})=Y\rho(\bm{x})\,,
\end{equation}
where $\Delta$ is the Laplace-Beltrami operator on the torus and
$\rho(\bm{x})$ is the total topological charge density
\begin{equation}\label{eq:charge_density}
\rho(\bm{x})=\frac{\pi}{3}\sum_{k=1}^{N}q_{k}\delta(\bm{x},\bm{x}_{k})-K(\bm{x})
\end{equation}
of $N$ disclinations located at the sites $\bm{x}_{k}$ plus a
screening contribution due to the Gaussian curvature $K(\bm{x})$ of
the embedding manifold. The first term in Eq. \eqref{eq:free_energy}
represents the long-range elastic distortion due to defects and
curvature. Its form resembles the potential energy of a system of
electrical charges. In this analogy the Gaussian curvature
$K(\bm{x})$ plays the role of a non-uniform background charge
distribution while defects appear as positively and negatively
charged point-like particles. As a result, disclinations arrange
themselves so to approximately match the Gaussian curvature. The
second term in Eq. \eqref{eq:free_energy} is the defect core-energy
representing the energy required to create a single disclination
defect. This quantity is related to the short-distance cut-off of
the elastic theory and is proportional to the square of the
topological charge times a constant $\epsilon_{c}$. Finally $F_{0}$
is an offset corresponding to the free energy of a flat defect-free
monolayer. 

\begin{figure}[t]
\centering
\includegraphics[width=.99\columnwidth]{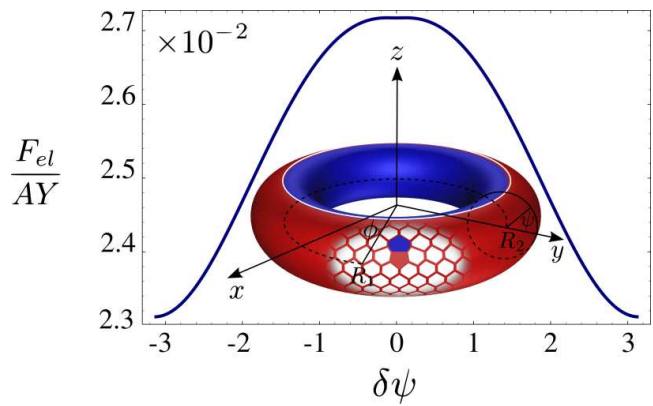}
\caption{\label{fig:redbluetorus}(Color online) Elastic energy of a 
$5-7$ disclination dipole constrained to lie on the same meridian, 
as a function of the angular separation. In the inset, illustration 
of a circular torus of radii $R_{1}>R_{2}$. Regions of positive and
negative Gaussian curvature have been shaded in red and blue
respectively. A standard parametrization of the torus is obtained by
considering the angles $\psi$ and $\phi$.}
\end{figure}

Using standard analysis the function
$\Gamma(\bm{x})$ can be written in the form
\begin{equation}
\Gamma(\bm{x})=\frac{\pi}{3}\sum_{k=1}^{N}q_{k}\Gamma_{d}(\bm{x},\bm{x}_{k})-\Gamma_{s}(\bm{x})\,,
\end{equation}
where $\Gamma_{s}(\bm{x})$ represents the stress field due to the
curvature of the torus and is given by:
\begin{equation}\label{eq:gamma_screening}
\frac{\Gamma_{s}(\bm{x})}{Y} =
\log\left[\frac{r+\sqrt{r^{2}-1}}{2(r+\cos\psi)}\right]+\frac{r-\sqrt{r^{2}-1}}{r}
\ ,
\end{equation}
where $r=R_{1}/R_{2}$ is the aspect ratio of the torus. The function
$\Gamma_{d}(\bm{x},\bm{x}_{k})$ is the stress field at the point
$\bm{x}$ arising from the elastic distortion due to a defect at
$\bm{x}_{k}$ and is given by
\begin{gather}
\frac{\Gamma_{d}(\bm{x},\bm{x}_{k})}{Y}
= \frac{\kappa}{16\pi^{2}}\left(\psi_{k}-\frac{2}{\kappa}\,\xi_{k}\right)^{2}-\frac{1}{4\pi^{2}\kappa}(\phi-\phi_{k})^{2}\notag\\[5pt]
+ \frac{1}{4\pi^{2}r}\log(r+\cos\psi_{k})-\frac{\kappa}{4\pi^{2}}\Real\{\Li(\alpha e^{i\psi_{k}})\}\notag\\[7pt]
+
\frac{1}{2\pi}\log\left|\vartheta_{1}\left(\frac{z-z_{k}}{\kappa}\bigg|\frac{2i}{\kappa}\right)\right|\,,
\label{eq:gamma_defects}
\end{gather}
where $\xi=\kappa\tan^{-1}(\omega\tan\psi/2)$ is a conformal angle
arising from the mapping of the torus to the periodic plane,
$z=\xi+i\phi$, and $\alpha$, $\kappa$ and $\omega$ are dimensionless
constants depending on the aspect ratio $r$:
\begin{gather*}
\alpha = \sqrt{r^{2}-1}-r\,,\qquad
\kappa = \frac{2}{\sqrt{r^{2}-1}}\,,\qquad
\omega = \sqrt{\frac{r-1}{r+1}}\,.
\end{gather*}
Finally $\vartheta_{1}$ is a Jacobian theta function and reflects
the double periodicity of the torus \cite{Theta}.

\begin{figure}[t]
\centering
\includegraphics[width=1\columnwidth]{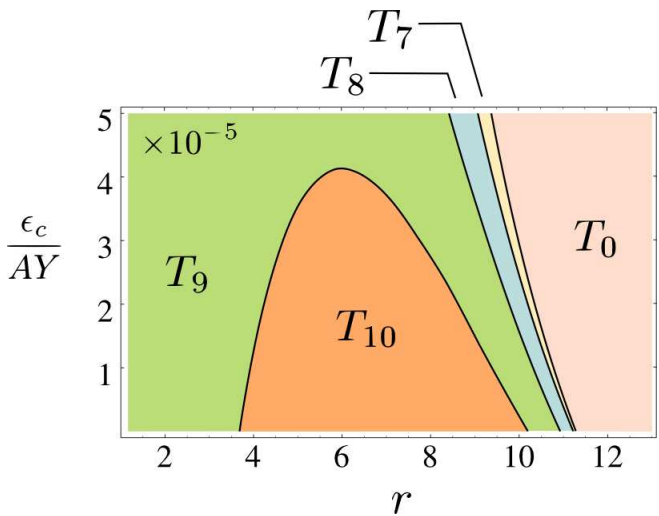}
\caption{\label{fig:phase_diagram1}(Color online) Phase diagram for 
$T_{p}$ configurations in the plane $(r,\epsilon_{c}/AY)$. For
$r\in[3.68,\,10.12]$ and $\epsilon_{c}\sim 0$ the structure is 
given by a $T_{10}$ configuration with symmetry group $D_{5h}$.}
\end{figure}

To analyze the elastic energy Eq. \eqref{eq:free_energy} we first
consider the simplest case of two opposite sign disclinations lying
on the same meridian of a torus of area $A=4\pi^{2}R_{1}R_{2}$. The
elastic free energy of this system is shown in  Fig. \ref{fig:redbluetorus}
as a function of the angular separation between the two disclinations.
The minimum is obtained for the positive disclination along the external
equator of the torus (where $K(\bm{x})$ is maximally positive)
and the negative disclination along the internal equator (where $K(\bm{x})$
is maximally negative). The picture emerging from this
simple case suggests a procedure to systematically construct optimal
defect patterns for an arbitrary number of disclination pairs, by
placing the same number of equally spaced $+1$ and $-1$
disclinations along the internal and external equators respectively.
We name this configuration with the symbol $T_{p}$, where $p$ stands
for the total number of disclination dipoles.

A comparison between the free energy of different $T_{p}$
configurations as a function of aspect ratio and core energy is
summarized in the phase diagram of Fig. \ref{fig:phase_diagram1}. We
stress here that only $T_{p}$ graphs with $p$ even have an embedding
on the torus corresponding to lattices of the TP$\frac{p}{2}$ class.
Nevertheless a comparison with $p-$odd configurations can provide
additional information regarding the stability of $p-$even lattices.
The defect core energy entering Eq. \eqref{eq:free_energy} is equal
to $2p\epsilon_{c}$. Although dependent on the microscopic details
of the system, the constant $\epsilon_{c}/AY\sim 10^{-5}$ for a
crystal of roughly $10^{3}$ atoms. In the range $r\in[3.68,\,10.12]$
and $\epsilon_{c}\sim 0$, the structure is dominated by the $T_{10}$
phase corresponding to a double ring of $+1$ and $-1$ disclinations
distributed along the external and internal equators of the torus at
the vertices of a regular decagon. The corresponding lattice has
$D_{5h}$ symmetry group.

That this structure might represent a stable configuration for
polygonal carbon toroids has been conjectured by the authors of
Ref. \onlinecite{LambinEtAl:1995}, based on the argument that the 
36$^{\circ}$ angle arising from the insertion of ten pentagonal-heptagonal 
pairs into the lattice would optimize the geometry of a nanotorus
consistently with the structure of the $sp^{2}$ bonds of the carbon
network (unlike the 30$^{\circ}$ angle of the $6-$fold symmetric
configuration originally proposed by Dunlap). In later molecular
dynamics simulations, Han \cite{Han:1997} found that a $5-$fold symmetric lattice,
such as the one obtained from a (9,0)/(5,5) junction, is in fact
stable for toroids with aspect ratio less then $r\sim 10$. The stability, 
in this case, results from the strain energy per atom being smaller than 
the binding energy of carbon atoms. We have shown here, from first principles, 
that a $5-$fold symmetric lattice indeed constitutes a minimum of the
elastic energy for a broad range of aspect ratios and defect core
energies.

\begin{figure}[b]
\centering
\includegraphics[width=1\columnwidth]{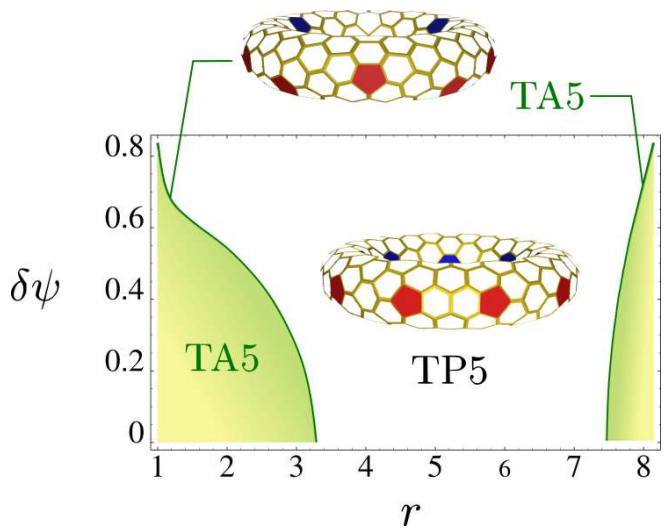}
\caption{\label{fig:phase_diagram2}(Color online) Phase diagram of a 
$5-$fold symmetric lattice in the plane $(r,\delta\psi)$. For small
$\delta\psi$ and $r$ in the range $[3.3,\,7.5]$ the prismatic TP5
configuration is energetically favored. For $r<3.3$, the system
undergoes a structural transition to the antiprismatic phase TA5.}
\end{figure}

For small aspect ratios the $5-$fold symmetric configuration becomes
unstable and is replaced by the $T_{9}$ phase. This configuration,
however, does not correspond to a possible triangulation of the
torus. In this regime, we expect the minimal energy structure to
consist of ten non-coplanar disclination pairs as in the
antiprismatic TA5 lattice. The latter can be analyzed by introducing
a further degree of freedom, $\delta\psi$, representing the angular
displacement of defects from the equatorial plane. A comparison
between the TP5 and TA5 configurations is shown in Fig.
\ref{fig:phase_diagram2} for different values of $\delta\psi$. For
small $\delta\psi$ and $r\in[3.3,\,7.5]$ the prismatic TP5
configuration is energetically favored. For $r<3.3$, however, the
lattice undergoes a structural transition to the TA5 phase.
For $r>7.5$ the prismatic symmetry of the TP5 configuration breaks
down again. In this regime, however, the elastic energy of both
configurations rapidly becomes higher because of the lower curvature
and defects disappear.

In the regime of large particle numbers, the amount of curvature
required to screen the stress field of an isolated disclination in
units of lattice spacing becomes too large and disclinations are
unstable to grain boundary ``scars'' consisting of a linear array of
tightly bound $5-7$ pairs radiating from an unpaired disclination
\cite{BowickNelsonTravesset:2000,GiomiBowick:2007}. In a manifold
with variable Gaussian curvature this process yields the coexistence
of isolated disclinations (in regions of large curvature) and scars.
In the case of the torus the Gaussian curvature in the interior is
always larger in magnitude than that in the exterior for any
$R_{1}>R_{2}$ and thus we may expect a regime in which the negative
internal curvature is still large enough to support the existence of
isolated $7-$fold disclinations, while on the exterior of the torus
disclinations are delocalized in the form of positively charged grain
boundary scars.

To check this hypothesis we compare the energy of the TP5 lattice
previously described with that of ``scarred'' configurations
obtained by decorating the original toroid in such a way that each
pentagonal disclination on the external equator is replaced by a
$5-7-5$ scar. The result of this comparison is summarized in the
phase diagram of Fig. \ref{fig:phase_diagram3} in terms of $r$ and
the number of vertices of the triangular lattice $V$ (the
corresponding hexagonal lattice has twice the number of vertices,
i.e. $V_{hex}=2V$). $V$ can be derived from the angular separation
of neighboring disclinations in the same scar by approximating $V
\approx A/A_{V}$, with $A_{V}=\frac{\sqrt{3}}{2}a^{2}$ the area of a
hexagonal Voronoi cell and $a$ the lattice spacing. When the aspect
ratio is increased from 1 to 6.8 the range of the curvature
screening becomes shorter and the number of atoms required to
destroy the stability of the TP5 lattice decreases. For $r>6.8$,
however, the geodesic distance between the two equators of the torus
becomes too small and the repulsion between like-sign defects takes
over. Thus the trend is inverted.

 On the material science side, defects can be functionalized
to provide binding sites for ligands. The number of excess
disclinations will then determine the valency of the surface, which
itself can serve as a building block for new molecules and bulk
materials \cite{Nelson:2002}. The universality of the predictions
presented here means that one has precise control of the valency by
tuning the aspect ratio of the torus. This could lead to a very
efficient scheme for creating a variety of basic surfaces with
well-defined valency which subsequently self-assemble into novel
molecules and bulk structures.

We acknowledge David Nelson and Alex Travesset for
stimulating discussions. This work was supported by the NSF through
Grant No. DMR-0219292 (ITR) and by an allocation through the
TeraGrid Advanced Support Program.

\end{document}